\documentclass[Afour,sageh,times]{sagej}

\usepackage{moreverb,url}

\usepackage[colorlinks,bookmarksopen,bookmarksnumbered,citecolor=red,urlcolor=red]{hyperref}

\newcommand\BibTeX{{\rmfamily B\kern-.05em \textsc{i\kern-.025em b}\kern-.08em
T\kern-.1667em\lower.7ex\hbox{E}\kern-.125emX}}

\def\volumeyear{2021}

\usepackage{textcomp}                  
\usepackage{mathptmx}                  
\usepackage{tabu}                      
\usepackage{booktabs}                  
\usepackage{multirow}
\usepackage{tabularx}
\usepackage{float}
\usepackage{multicol}
\usepackage[inline]{enumitem}
\usepackage[usenames,dvipsnames]{xcolor}
\usepackage[normalem]{ulem}
\definecolor{mred}{rgb}{.80,.12,.30}
\definecolor{grey}{rgb}{0.5,0.5,0.5}
\definecolor{outlinecolor}{rgb}{0,0.18,0.38}
\definecolor{Purple}{rgb}{.75,0,.85}
\definecolor{green}{rgb}{.11,.68,.15}
\usepackage{amsfonts}
\usepackage{amsmath}
\newif\ifnotes
\notesfalse

\newcommand{\zhenge}[1]{\ifnotes{\textcolor{red}{(Zhenge: #1)}}\fi}

\definecolor{mattblack}{RGB}{0,0,0}
\definecolor{mattorange}{RGB}{178,101,24}
\newcommand{\mbedit}[1]{\textcolor{mattblack}{#1}}

\newcommand{\eat}[1]{}

\definecolor{joshgreen}{RGB}{23,180,55}
\definecolor{joshpurple}{RGB}{178,24,101}

\definecolor{zhengered}{RGB}{255,0,0}

\newcommand{\myparagraph}[1]{\noindent \textbf{#1}\ }

\begin{document}

\runninghead{Zhao and Motta \textit{et~al.}}

\title{STFT-LDA: An Algorithm to Facilitate the Visual Analysis of Building Seismic Responses}

\author{}
\author{Zhenge~Zhao\affilnum{1}, Danilo~Motta\affilnum{2}, Matthew~Berger\affilnum{3}, Joshua A.~Levine\affilnum{1},\\
	Ismail~B.~Kuzucu\affilnum{4}, Robert~B.~Fleischman\affilnum{4}, Afonso~Paiva\affilnum{2}, Carlos~Scheidegger\affilnum{1}
}

\affiliation{\affilnum{1}the Department of Computer Science at the University of Arizona\\
	\affilnum{2}the University of S\~{a}o Paulo\\
	\affilnum{3}Vanderbilt University\\
	\affilnum{4}the Department of Civil Engineering at the University of Arizona
	}

\corrauth{Zhenge Zhao, the Department of Computer Science at the University of Arizona,
HDC Lab, Gould-Simpson Building, 1040 E. 4th Street, Tucson, Arizona, USA.}

\email{zhengezhao@email.arizona.edu}

\begin{abstract}
Civil engineers use numerical simulations of a building's responses to seismic forces to understand the nature of building failures, the limitations of building codes, and how to determine the latter to prevent the former. 
Such simulations generate large ensembles of multivariate, multiattribute time series.
Comprehensive understanding of this data requires techniques that support the multivariate nature of the time series and can compare behaviors that are both periodic and non-periodic across multiple time scales and multiple time series themselves.
In this paper, we present a novel technique to extract such patterns from time series generated from simulations of seismic responses. The core of our approach is the use of topic modeling, where topics correspond to interpretable and discriminative features of the earthquakes. We transform the raw time series data into a time series of topics, and use this visual summary to compare temporal patterns in earthquakes, query earthquakes via the topics
across arbitrary time scales, and enable details on demand by linking the topic visualization with the original earthquake data.
We show, through a surrogate task and an expert study, that this technique allows analysts to more easily identify recurring patterns in such time series. By integrating this technique in a prototype system, we show how it enables novel forms of visual interaction.
\end{abstract}


\maketitle

\section{Introduction}\label{sec:introduction}
\begin{figure*}
	\centering
	\includegraphics[width=\linewidth]{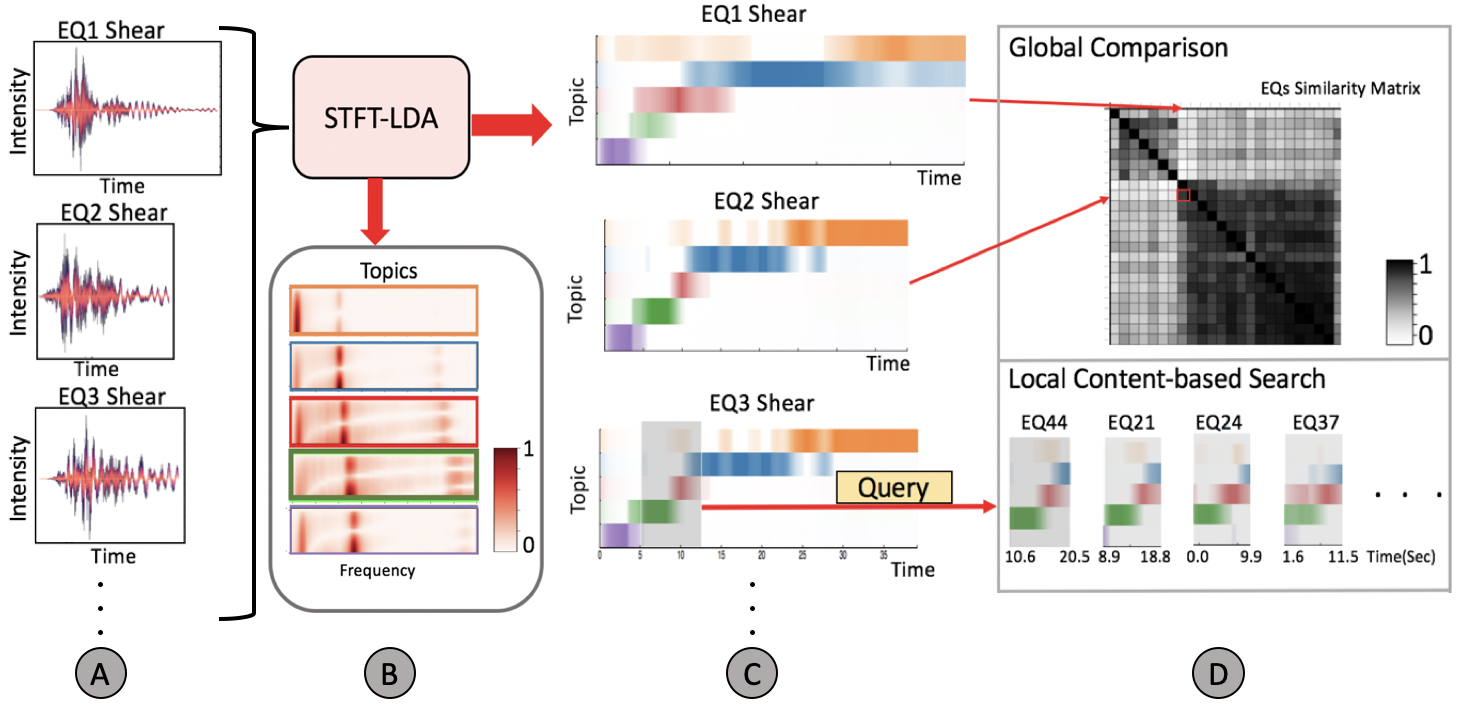}
	\caption{The technique we propose in this paper, STFT-LDA, captures variation patterns across multiple time and frequency scales, as well as different attributes in a multivariate time series (A). We show through one quantitative user studies and one expert evaluation (Section~\ref{sec:evaluation}) that the visual summaries (C) provide better discrimination of the behavior of such multivariate time series. STFT-LDA uses \emph{topic modeling} to capture variation patterns in the time series; in Section~\ref{sec:expert-user-study}, we show that the topics themselves in (B) are meaningful and interpretable. Finally, the information generated by STFT-LDA itself enables powerful visual interaction modalities (D). Section~\ref{sec:other-views} shows how the features enable a global overview that highlights overall similarities between the behaviors of all simulations, and Section~\ref{sec:content-based-search} shows how our technique supports different forms of visual interaction such as content-based search for visual filtering, and global comparison of an ensemble of earthquakes.\label{fig:teaser} }
\end{figure*}

In what ways do building structures fail during an earthquake?  This
question has serious legal and economic implications: building codes
such as the International Building Code~\citep{ibc2018} dictate safety
standards, and these can have an impact on how much buildings cost.
Building codes, in addition, do not necessarily reflect accurately the
complicated ways in which buildings sway and break, and thus are
constantly being revised since their introduction in the
1980s~\citep{kurama2018seismic,security2013risk}.  In this paper, we
present a novel technique for visually exploring data generated from
simulations of building responses under seismic loads, and a prototype
system built to support the visualization of such data.

Civil engineers often use small-scale, real-life experiments to
understand the dynamics of buildings during such
earthquakes~\citep{schoettler2009preliminary}.  This process can be slow
and laborious, but the data analysis is comparatively straightforward.
More recently, there has been a tendency to use simulated
shake tables to explore a variety of scenarios and to study the forces
and stresses exerted on buildings during such earthquakes.  One of the
central advantages of computational science is the drastic reduction in
experimental costs: studies that once were prohibitively expensive and
laborious to run are now performed entirely \emph{in silico}. As a
consequence, civil engineers now have an over-abundance of data, and the
barrier to the creation of safer building and building codes is no
longer in the creation of such studies, but rather in understanding the
results of such simulations. In such a scenario, data analysis and
interactive visualization play a critical role.

Specifically, the data generated by such computational simulations is
a large ensemble of multivariate time series. These simulations take
as input a specification of the structure of the building, and the
ground acceleration of a recorded earthquake. Each run of the simulation 
records a number of physical variables (such as displacement, shear, moment, and acceleration) at a relatively high frequency (typically 400 samples per simulation second). Each of these variables is
recorded for each of the building floors' degrees of freedom for the duration of the
earthquake.

Engineers wish to understand and compare the
responses across a number of different earthquakes.
The resulting ensemble of time series data has both
periodic and non-periodic components. As we will show in
Section~\ref{sec:technique}, the frequency components change
throughout the earthquake, which means that traditional
frequency-domain analysis is not particularly well-suited. In
response, in this work we collaborated with civil engineers that produce and study such data, in order to design a visualization that helps them in their analysis. Specifically, we contribute:

\begin{itemize}
	\item a novel technique to extract periodic and non-periodic features from
	time-series ensembles that combines short-time Fourier Transforms
	with topic modeling (Section~\ref{sec:technique}),
	\item a coordinated multiple-view prototype system that leverages the advantages of visually encoding topics and incorporates both local and global features (Section~\ref{sec:system}), and 
	\item one quantitative study and an expert evaluation which show that this technique can provide a better infrastructure for visual analysis of this specific
	type of time-series data (Section~\ref{sec:evaluation}).
	
\end{itemize}

\section{Related Work}

Our visualization approach builds on concepts from time series analysis and visualization of topic models.  We also describe recent research in the visualization of seismic data.

\subsection{Visualizing Time Series Data}

Data having a temporal component appears frequently in a variety of
settings, and thus there are numerous works on visualizing time series
data.  We highlight those works that study multivariate time series.
We refer the reader to Aigner et al.~for a more complete overview of time series visualization~\citep{aigner2011visualization}.

Some of the earliest work on time series visualization focuses on questions of layout and best design choices for the display of time series data.  Keim et al.~align pixel dense visualizations of time series to high recursive patterns~\citep{keim1995recursive}.  Weber et al.~use a spiral metaphor to draw time series data, interleaving multiple spirals when the data is multivariate~\citep{weber2001visualizing}.  Carlis and Konstan also use spirals, but stack multiple variables in 3D~\citep{carlis1998interactive}.  Byron and Wattenberg use streamgraphs to visualize multivariate time series data~\citep{byron2008stacked}.  More recent work has focused on how best to interact with time series.  The TimeSearcher tool of Hochheiser and Shneiderman provides an approach to interactively perform range queries of multiple time series data~\citep{hochheiser2003interactive}, which was later extended by Buono et al.~to provide example-based querying~\citep{buono2005interactive}.  The LiveRAC system of McLachlan et al.~couples multiple views and semantic zooming to present visualizations of system management data~\citep{mclachlan2008liverac}.

Our work is most related to visual analytics systems that can help users identify patterns and structure within time series.  Buono et al.~use similarity-based forecasting to search for patterns in historical time-series data and visualize predictions of future behavior~\citep{buono2007similarity}. Guo et al.~use their EventThread system to cluster event sequences into categories using tensor analysis~\citep{guo2018eventthread}.  Lin et al.~decompose time series using symbolic aggregate approximation to construct a hierarchical representation of patterns generated by a sliding window~\citep{lin2004visually}.  Others have studied designing user-driven contexts based on novel interactions.  In particular,  Muthumanickam et al.~explore long time series by constructing a grammar of basic shapes based on user sketches~\citep{muthumanickam2016shape}.  Correll and Gleicher also study sketching for time series~\citep{CG16}.

Finally, J{\"a}ckle et al.~explore multivariate time series data by constructing 1D MDS plots over sliding temporal windows~\citep{jackle2016temporal}.  We similarly use a sliding window in the STFT, but our windowing scheme is designed to capture how frequency usage evolves over time.  Miranda et al.~construct a ``pulse`` to identify cyclic patterns in time-series based on urban data~\citep{miranda2017urban}.  Instead of identifying cyclic patterns with topological tools, we focus on periodic structures that exist at various frequencies.  The structures translate to important features and visualization primitives that convey intuition about the frequency domain.

\subsection{Topic Modeling and Visualization}

Topic modeling has been utilized frequently in the context of visualizing text data.  In particular, our work utilizes latent Dirichlet allocation (LDA), pioneered by Blei et al.~\citep{Blei:2003:LDA:944919.944937} as a probability generalization of latent semantic analysis~\citep{landauer1998introduction}.

In the field of text visualization, topic modeling is frequently used as a data processing step to provide more meaningful structure to unorganized documents.  One feature of topic models is that they offer a means to project individual documents into a lower dimensional (typically 2D) spaces, providing views of the latent structure~\citep{herr2009nih,iwata2008probabilistic}.  Termite relies on an alternative display of topic models, using a tabular view that helps a user understand the distributions of terms both within and across topics~\citep{chuang2012termite}.  Dou et al.~user the parallel coordinates metaphor to display LDA models in the ParallelTopics system~\citep{dou2011paralleltopics}.  The UTOPIAN system of Choo et al.~uses force-directed layout to display topic models~\citep{ChooLRP13} and Lee et al.'s~iVisClustering system couples graph layouts with other views to interactively steer the LDA model~\citep{lee2012ivisclustering}.  Providing supervision to LDA has beed studied by El-Assady et al.~\citep{el2018progressive} who provide an iterative framework to adjust topics, informed by user studies by Lee et al.~\citep{lee2017human}.  Alexander and Gleicher use buddy plots to visually compare multiple topic models and to derive comparison and understanding tasks~\citep{alexander2016task}.

More closely related to our own work are approaches that visualize connections between topic models and time. Luo et al.~couple event-based analysis for text collections to identify topics in a time series view in EventRiver~\citep{luo2012eventriver}.  Wei et al.'s~TIARA visualizes the evolution of topics over time~\citep{wei2010tiara}, by using a modified ThemeRiver~\citep{havre2000themeriver} display.  Cui et al.~also employ the metaphor of rivers, but focus on visualizing where specific events happen in the evolution of topics in TextFlow~\citep{cui2011textflow}.  The approach of Leadline is to associate topic themes with specific events, highlighting topic streams by their length and burst behavior~\citep{dou2012leadline}.  

While most works that couple topic modeling with time focus on visualizing document collections, LDA extends beyond just documents.  Chu et al.~use LDA to discover topics from taxi trajectories~\citep{chu2014visualizing}.  Hong et al.~use LDA to explore unsteady flow, mapping flow features correspond to words~\citep{hong2014flda}. Chen et al.~use LDA to categorize operation behaviors in the security management system ~\cite{8663312}. All these works share similarity to our own in that abstract, data-dependent concepts map to traditional components in topic modeling.

\subsection{Frequency-domain Analysis}
\mbedit{The frequency-domain representation of time series data is a useful analysis tool for seismologists, as they are often interested in understanding periodic and non-periodic features. Fourier-based decomposition is widely used for filtering noise and helping to identify periodic phenomena~\citep{Singh20160871,doi:10.1190/geo2015-0489.1,09a002063d5645d9aea0ce91cd946194}, yet for time-localized behavior, Fourier methods are unsuitable. Short-Time Fourier Transform (STFT) preserve the frequency content dynamics over time, by shifting a spatially-compact window and calculating the Fourier transform in each small window~\citep{allen1977short}. Wavelet Analysis~\citep{Chakraborty2000FrequencytimeDO} is similar to STFT, but instead of using a fixed window, Wavelets enable multiresolution analysis via a set of windows whose functions resemble tiny waves that grow and decay in spatial support. Within seismology, Sinha et al.~\citep{doi:10.1190/1.2127113} employ the Continuous Wavelet Transform (CWT), Wang et al.~\citep{6809966} apply the Synchrosqueezing Transform to the seismic signal to achieve a higher precision than CWT, and Wang et al.~\citep{doi:10.1190/1.2387109} uses Matching Pursuit Decomposition to automatically determine the best spatial resolution.}
\eat{
	The frequency-domain analysis of a time series could provide extra features which are difficult to demonstrate in the time domain. Seismologists analyze seismic responses in the frequency domain trying to understand periodic and non-periodic features. There exists many types of time-frequency analysis methods. Fourier based decomposition is widely used for extracting specific behaviors and filtering noise in the responses~\citep{Singh20160871,doi:10.1190/geo2015-0489.1}. Short-Time Fourier Transform(STFT) preserve the frequency content changes over time, by shifting a time window and calculating the Fourier transform in that small window~\citep{allen1977short}. Wavelet Analysis~\citep{Chakraborty2000FrequencytimeDO} is similar to STFT, instead of using a fixed window, it unitizes a set of windows whose functions resemble tiny waves that grow and decay in short period of time. Sinha et al.~use Continuous Wavelet(CWT) Transform to do spectral decomposition of seismic data~\citep{doi:10.1190/1.2127113}. Wang et al.~\citep{6809966} apply Synchrosquesszing Transform to the seismic signal to achieve a higher precision than CWT.  Wang et al.~\citep{doi:10.1190/1.2387109} proposes Matching Pursuit Decomposition automatically returning the best resolution. 
}

\mbedit{While these methods provide useful analysis tools for understanding the time-frequency representation of a single, or few, time series, they are poorly suited for handling the large amounts of time series that the domain experts we work with typically face. Visually analyzing a large amount of spectrograms -- beit produced from STFT or Wavelets -- is cognitively demanding, and few methods exist that can help summarize such data. For instance, simply taking an average of spectrograms might obscure important details, while dimensionality reduction techniques fail to retain the time-frequency representations that are of interest to the domain experts.}

\eat{
	While these methods provide spectrogram tools for time-frequency representation, they have some common deficiencies. First, all these spectrograms they create is not simple to interpret, the frequencies distributions are originally shown to the domain experts without summarizing and abstracting. Therefore, comparing multiple responses and quickly identifying patterns are still difficult. Moreover,
	general techniques such as taking average or simply applying some dimension reduction techniques like principal component analysis (PCA) will lose information like different behaviors of floors and attributes.
}

\subsection{Visualization of Seismic Data}
We also discuss other efforts in visualization that focus on visualizing seismic and earthquake data.  These techniques usually emphasize two- and three-dimensional views, typically coming from either simulation and/or observational data.  Much of this research has focused on techniques, such as using images~\citep{akcelik2003high}, video~\citep{chourasia2008visualizing}, or volume rendering~\citep{ma2003visualizing, yu2004parallel} to display earthquake data.  Chopra et al.~deploy an immersive virtual environment to visualize earthquake simulations for domain scientists~\citep{chopra2002immersive}.  Wolfe et al.~use ultrasound reflection to visualize seismic simulations as volume data~\citep{wolfe1988interactive}.  While these techniques are powerful, we note that they focus on significantly different data than what we present in this work, as we visualize a building response to a measured earthquake instead of the earthquake itself.

Even using different data, visualization systems for earthquake visualization are motivational to the analysis we employ, in particular in how they couple simulation with measured data.  Yuan et al.~present a complete visual system for studying earthquake data from multimodal sources, including measured data~\citep{yuan2010scalable}.  Patel et al.~interpret measured seismic data using the Seismic Analyzer to illustrate 2D seismic data~\citep{patel2008seismic}.  Hsieh et al.~also visualize time-varying field-measured data to produce time-varying volumetric renderings~\citep{hsieh2010visualizing}.   Komatitsch et al.~provide comparisons of seismic waveforms produced between simulation and observational data~\citep{komatitsch200314}.  We again emphasize that while these works are inspirational in terms of studying seismic response, our work differs significantly in that we are focused on trends that help us compare numerical simulations of buildings under seismic loads.

\section{Problem Setup}
\label{sec:probset}

We first describe the data that the civil engineers tend to produce via numerical simulation.
In studying the behavior of buildings that experience earthquakes, civil engineers are concerned with analyzing simulations that produce sets of multivariate time series.
A single simulation produces a time series that expresses each physical variable on each degrees of freedom of a floor for given an input, recorded earthquake signal.
In this study, the simulations track 6 physical attributes of interest, for each floor: acceleration, shear, diaphragm force, moment, drift ratio, and interstory drift ratio. Each physical attribute is an important indicator of the building status. For example, shear measures the cumulative force parallel to each floor, while interstory drift ratio measures the positional difference between two floors at a given point in time. This gives a vector-valued time series for each attribute, and each simulation has 25,000 time steps in average in this study. Each attribute is normalized by dividing the raw value by a predetermined design limit. This has the benefit that any value of
the time series above 1 or below negative 1 indicate that the building is operating out of its safe design specifications, and mitigates the
issue of comparing variables of different units. For simplicity of discussion, in what follows we treat the combinations of floors and variables as a combined multivariate signal (and thus, we abuse language at times and simply refer to the quantity of interest as \emph{floor} or \emph{variable}).

\subsection{Glossary of Seismological Terms}
\myparagraph{Earthquake Simulation} A vibrational input that possesses the essential features of a real seismic event is applied to structures to study the effects of earthquakes on structures. (Section ~\ref{sec:introduction})

\myparagraph{Story Shear} A term to measure the force parallel to each ﬂoor in a building. (Section ~\ref{sec:preliminary-design-study})

\myparagraph{Mode} One of a set of independent vibration conﬁgurations a building can exhibit. Higher modes correspond to more complex conﬁgurations. Buildings can vibrate in multiple modes simultaneously. (Section ~\ref{sec:preliminary-design-study})

\myparagraph{Ground Acceleration} A time-varying attribute of the earthquake directly indicating the acceleration of the ground at a particular point in time. (Section ~\ref{sec:details-views})

\myparagraph{Impulse} A term describing a force applied onto the building by the earthquake over a very short period of time. (Section ~\ref{sec:expert-user-study})

\myparagraph{Elastic and inelastic state} Under an elastic vibration, the building can resume to undeformed initial position when the external force is removed; however, inelastic vibration causes irreversible damage to the structure, and it may remain deformed even after the removal of the external force. (Section~\ref{sec:expert-user-study})

\begin{figure}[t]
	\centering 
	\includegraphics[width=\columnwidth]{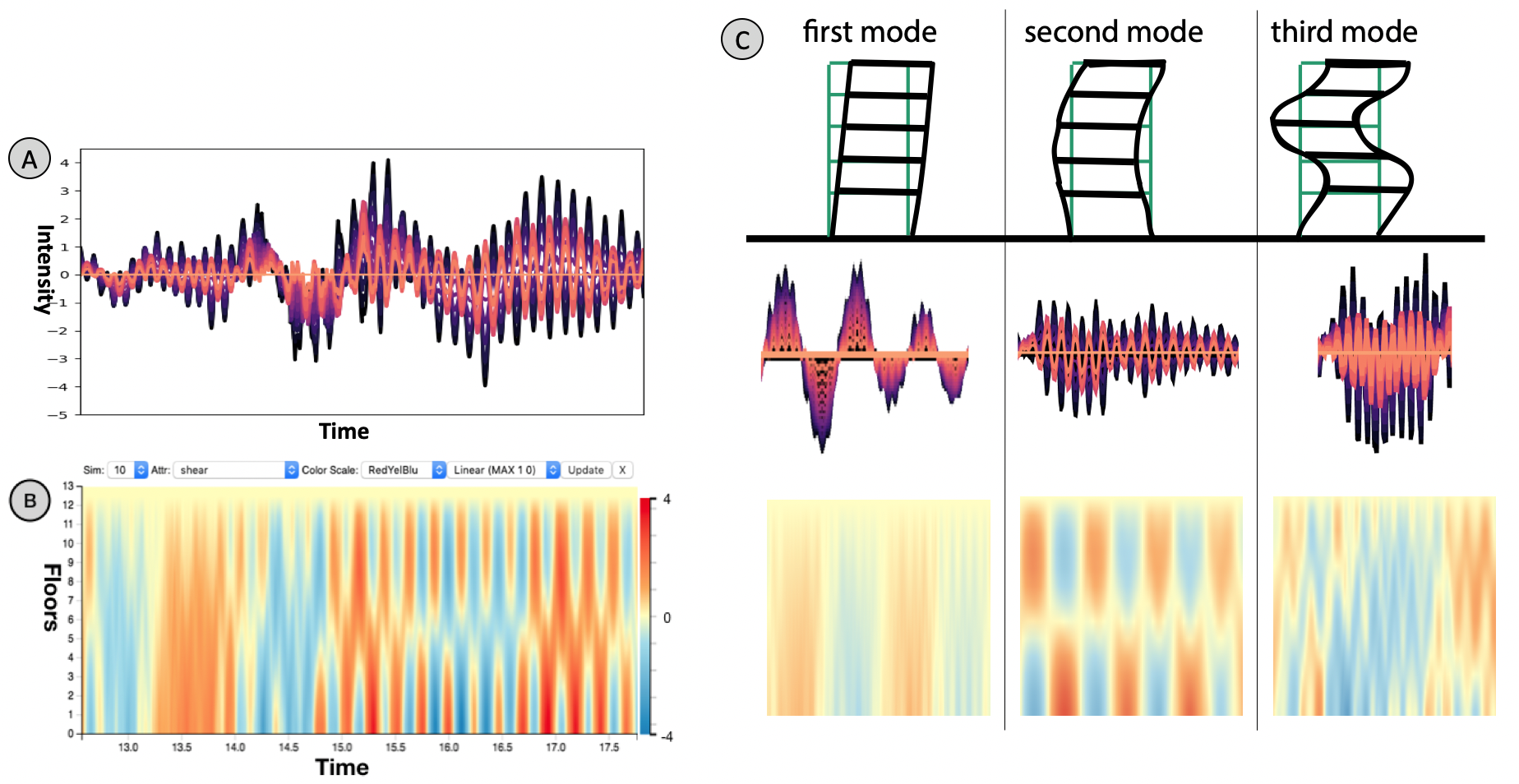} 
	\caption{The preliminary system utilizes a 2D heatmap (B) to visualize a time series across different floors (A). The variation of the color indicates the change of one physical attribute, which also reflects the vibration condition of the building. A linear interpolation is applied to the values between floors to facilitate comparisons across floors and different timestamps. The color encoding simplifies the recognition of three basic vibration modes in (C), nevertheless, this method lacks the ability of summarizing the simulation behavior as well as making comparisons across different simulations. See Section~\ref{sec:technique} for the technique we propose to solve these problems. }
	\label{fig:preliminary-design-study}
\end{figure}

\subsection{Preliminary Design Study}
\label{sec:preliminary-design-study}
Through regular meetings with civil engineers, we found that a key aspect of their analysis is understanding response and the fundamental vibration modes of the building, often determined by the mass of stiffness of the building. 
In particular, studying these time series helps them understand the fundamental vibration periods of the building, often determined by the height, support structure, and materials used to construct the building.  As all objects have a natural vibrational period, understanding where deviations occur can often be indicative of damage. More specifically, the vibration behavior of a building can exist in different modes (Fig. \ref{fig:preliminary-design-study}(C)) where either all floors are vibrating in alignment or out of alignment. For example, if all floors move in the same direction, back and forth, the building will vibrate like a pendulum swing. The civil engineers typically refer to a building in this state as a \emph{first mode}. If the building undertakes shear stress from different directions at the same time, it will bend like an ``S'' shape, which is typically referred to as a \emph{second mode}. If the building is swinging in the shape of an ``M'' or ``W'', then this is typically called a \emph{third mode}. In seismic analysis, these motions have received long-lasting focuses and civil engineers conduct many experiments in order to understand the internal connections between the mode behaviors of a building and earthquake~\citep{article1,article2,KRAWINKLER1998452}.

Currently civil engineers use simple visualizations such as line plots (Fig.~\ref{fig:preliminary-design-study}(A)) to plot a building's response to individual earthquake. However, it can be complex to directly analyze line plots as the data is measured across a range of floors and variables. Quickly spotting the mode of a building is challenging in this scenario, as differences between modes manifest as subtle visual differences. Moreover, in a real-world scenario,  the building's motion in an earthquake is often more complicated than simply three \emph{modes}. Specifically, the movement is often disorderly and it also evolves slowly in response to damage, which leads to an evolution of material ductility that eventually alters the vibrational modes. 

\begin{figure}[t]
	\centering 
	\includegraphics[width=\columnwidth]{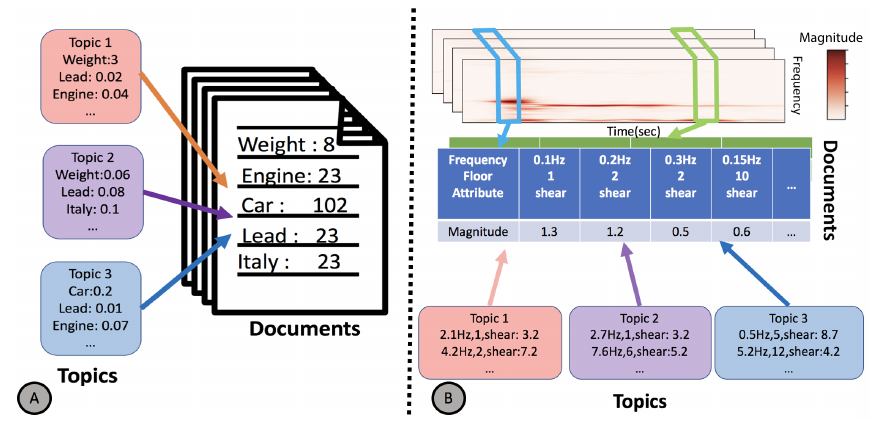} 
	\caption{Traditionally, LDA is used to summarize different
		distributions of word frequencies in documents into topics.
		In our paper, we use LDA to summarize distributions of
		frequency patterns obtained from STFT. Each ``document'' in
		our case is a collection of frequency distributions from each 
		of the different attributes on the multivariate time series
		(specifically, one time series for the shear strength measured
		in each floor).}
	\label{fig:LDA}
\end{figure}

Thus, for our initial task, we built an infrastructure to enable civil engineers to
visually explore multivariate time series and spot interesting patterns such as vibrational modes.
Towards this end, we built a prototype interface using 50 earthquake simulations provided by the engineers.
For each simulation, we utilize a 2D heatmap to visualize the response of each single physical variable plotted over time and building floor. 
As shown in Fig.~\ref{fig:preliminary-design-study}(B), users have access to different simulations and the corresponding physical attributes, and they can 
also choose different color scales and mapping methods to emphasize various patterns.
We showed this visualization tool to the civil engineers and they agreed that this view is supportive in spotting periodic behaviors as well as understanding the deviations across both floor and time steps directly.
In the meanwhile, choosing different color maps could help them simplify the signals and highlight interesting phenomena.
In particular, civil engineers can roughly observe the main vibration mode of the building by reading the color patterns. 
Taking the shear attribute of an earthquake for example, in Fig.~\ref{fig:preliminary-design-study}(C), 
if the building is in the \emph{first mode}, the colors of all the floors are either red
or blue at the same time.
On the other hand, if the building is in the second mode,
the colors are always different for the upper and lower floors at the same time steps, which reflects that the directions of the shear attribute for corresponding floors are also opposite.

In moving to studying multiple earthquake simulations, however, the 2D heatmap has limitations. First, even though it reduces the complexity of the 
origin multivariate time series, the visualization is still too complicated for users to understand general patterns and make comparisons across different simulations. 
For example, civil engineers may spot some of the mode behaviors in a simulation, but the user may still need to recall and match these color patterns back and forth while inspecting another simulation. Secondly, this direct visualization of time series doesn't help much in answering important questions like how the frequencies change throughout the earthquake or what is the highest intensity of the frequencies. Finally, this visualization is not scalable when more variables are introduced, specifically, considering two physical attributes at the same time. This is also a problem in previous methods when we are trying to visualize sets of frequency components across all earthquakes and their variables.

\subsection{Task Abstraction}
\label{sec:task-abstraction}
In visually exploring multiple earthquake simulations, 
there are a set of tasks that civil engineers wish to achieve:
\begin{itemize}
	\item \textbf{(T1) Summarizing Earthquake Behaviors.} Civil engineers would like to understand the space of discriminative earthquake behaviors.
	\item \textbf{(T2) Exploring Collections of Earthquakes.} It is challenging for civil engineers to even know where to begin their study. Having
	a general overview of earthquakes can help them decide what earthquake, or set of earthquakes, to study first.
	\item \textbf{(T3) Exploring Time-localized Earthquake Features.} Given a single earthquake, the civil engineers would like to understand
	how an identified feature at a specific time interval relates to other earthquakes.
	\item \textbf{(T4) Identifying Deviations and Outliers in the Set.} In addition to summarizing the aggregate behavior of earthquakes, the civil engineers also seek to understand which combinations of parameters/inputs produce results that deviate from the expected behavior.
\end{itemize}

\begin{figure}[t]
	\centering 
	\includegraphics[width=\columnwidth]{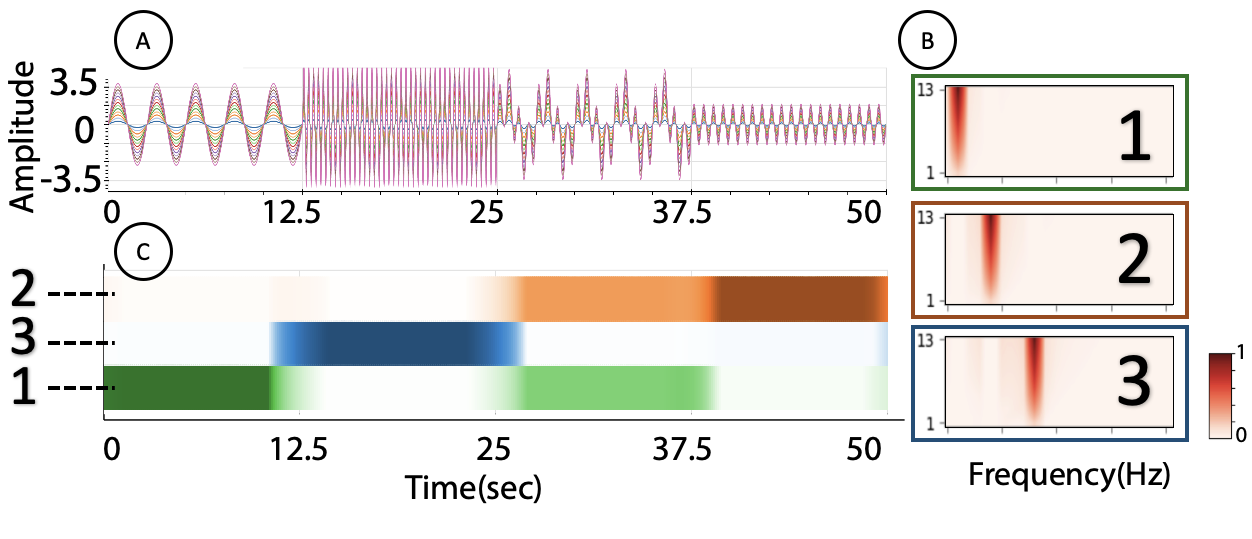} 
	\caption{A synthetic, multivariate time series generated to
		illustrate the behavior of STFT-LDA. The time series (A) goes
		through four different phases, characterized by different
		amplitudes and frequencies. In this case, we use STFT-LDA to
		generate three topics (B) which characterize the overall
		variability in the time series, and the summary view (C) shows
		how the patterns change over time and how they relate to one another.}
	\label{fig:SemanticData}
\end{figure}

Fundamental to satisfying these tasks is a notion of \emph{earthquake similarity}, taken with respect to arbitrary time intervals.
Similarity enables overviews of earthquake simulations, as measured across their entire duration, allowing the user to identify one or a small set of earthquakes
to begin their analysis \textbf{(T2)}.
Given a single earthquake, similarity also enables the user to query earthquakes, either globally or locally in time, allowing the user to compare
earthquakes at different time scales \textbf{(T3)}.  Finally, earthquakes that are dissimilar from the set can help to identify where large deviations have occurred that might necessitate further investigation \textbf{(T4)}.

While there are many ways one can compute similarity between the time series data produced by earthquake simulations, in this work we seek a mechanism to produce an \emph{interpretable} similarity measure.  Key to this interpretation is producing visual representations of earthquakes that enable civil engineers to understand why, when, and where earthquakes are similar.
Unfortunately, no technique in the literature supports such demands.
The core of our approach is a method to transform multivariate time series into such a representation that improves how users visually comprehend trends
and patterns across time series.
In particular, we model multivariate time series through \emph{topic modeling}.
Concretely, each multivariate measurement in time is replaced by a distribution of topics that best explain the earthquake at that point in time.

Our topic model is designed in such a way that each topic, viewed as a time series, smoothly changes over time, and thus it is far
easier to comprehend than the original earthquake simulation measurements.
Furthermore, each topic is characterized as a distribution over frequencies for each variable, and thus topics are interpretable with respect
to earthquake behaviors of interest to the civil engineers.
This enables civil engineers to comprehend general earthquake behaviors by inspecting the individual topics \textbf{(T1)} as a mechanism for summarizing the group.
These time-varying topic distributions underlie our visual analytics approach to exploring earthquake collections, as this representation drives
how we compute similarity between earthquakes.

\section{STFT-LDA: Topic Modeling for Multivariate Time Series}
\label{sec:technique}

At the core of our visual analytics technique is a novel representation of multivariate time series, designed to capture domain-specific features
in a manner that enables effective visual exploration.
Time series data produced from earthquake simulations can be characterized as having periodic behavior that varies over time, where changes in periodicity often reflect
different phases of the earthquake simulation.
To capture time-varying periodic phenomenon, we use Short-Time Fourier Transform (STFT)~\citep{allen1977short}, individually computed over all time series for each earthquake simulation.
Although descriptive of earthquake behavior, the STFT alone does not make it easier for the user to perform visual exploration of collections of multivariate time series data.
To this end, we perform topic modeling on the set of STFT, and use the learned topics for both visually encoding time series, as well as computing similarity over time series.
We discuss the STFT and topic modeling in more detail below.

\subsection{Short-time Fourier Transform (STFT)}
Besides inspecting the time series of seismic responses, civil engineers are concerned with their frequency domain representation, in order to distinguish periodic behaviors of
earthquakes and how periodic behavior changes as the earthquake simulation progresses.
The STFT is a suitable transformation for this purpose, as it captures the time-localized frequency content of a signal.
The STFT is constructed by defining a temporal window of fixed size, which we denote by $\tau$, sliding the window over the signal, and for each window the Fourier transform
is computed on the subset of the signal that resides within the window.
This results in a sequence of frequency decompositions, one for each window, for each time series of each variable in an earthquake.
Fig.~\ref{fig:LDA}(B) shows an example of the STFT applied to an earthquake time series of a single variable (shear) across thirteen floors.  The STFT is shown as a color mapped spectrogram, where the $x$-axis refers to time steps, the $y$-axis refers to frequency, and the color intensity refers to the squared frequency magnitude of the FFT.


\subsection{Topic Modeling STFTs using Latent Dirichlet Allocation (LDA)}
Although the STFT is descriptive of the phenomena present in the earthquake time series, it is not an ideal visual encoding for exploration.
It is necessary for the user to visualize the STFTs across all variables, but such views scale poorly in the number of variables.
To build a visual representation that compactly represents a set of STFTs, we turn to topic modeling.
Topic modeling has traditionally been used to obtain a better understanding of textual data.
More specifically, as illustrated in Fig.~\ref{fig:LDA}(A), given a set of documents where each document is comprised of a set of words and corresponding word counts,
\emph{topics} are learned from the data such that each topic is a mixture over words, and documents are mixtures over topics.
The topics are meant to capture latent themes in the document corpora, with each document typically represented with a few predominant topics, or themes, rather than
its original set of words.

In our scenario, we treat a multivariate time series earthquake as a \emph{time series of documents}.
More specifically, our vocabulary of \emph{words} corresponds to a binned set of frequencies.
In particular, we treat frequencies corresponding to different variables in the earthquake as being distinct, thus our vocabulary $W$ for $m$ uniformly discretized frequency ranges
and $n$ variables is of size $|W| = m \cdot n$. 
Each \emph{document} is formed at a given time by cascading all STFT windows over all variables, taking the document's word count as the frequency magnitude.
We then perform Latent Dirichlet Allocation (LDA)~\citep{Blei:2003:LDA:944919.944937} over documents that come from all of the earthquakes, which results in a set of topics.
We do not normalize documents because different earthquakes can have different frequency magnitudes, and we prefer the topic modeling to be sensitive
to these variations.
On the other hand, for each document LDA produces probabilities over topics, and thus the words over a topic, namely the weights over frequencies and variables,
cannot be interpreted as probabilities.
We thus normalize the topics, individually for each topic, allowing us to comprehend importance of words in a relative manner.
Each topic thus outputs a mixture of variable-dependent frequencies, as illustrated in Fig.~\ref{fig:LDA}(B).


\subsection{Illustrative Example}
\label{sec:illustrative}

We use the topics for visualization with two different views, see
Fig.~\ref{fig:SemanticData} for an illustration of a synthetic example modeled with three topics.  First, we visualize a given
multivariate signal by visually encoding each document in its time series
through its topic distribution, as shown in Fig.~\ref{fig:SemanticData}(C).  This view shows colored stripes to visualize each topic and its evolution across time.  
Each row is associated with a given topic, and we opacity-map each document's
normalized topic weight.  Within any given column, the topic weights will sum to $1.0$.

Second, we compactly visualize a topic as a 2D scalar field,
where the $x$-axis represents frequency, the $y$-axis represents variable, and
we color map the probability of each word belonging to the topic, as shown in
Fig.~\ref{fig:SemanticData}(B).  In this manner, the user can identify patterns
and transitions in the time-series document-topic view, and access
details-on-demand in the topic view.

We illustrate how these views help describe the original signal (Fig.~\ref{fig:SemanticData}(A)).
Our synthetic example models a single earthquake that consists of 13 time series where each time series has 4 phases, while the time series lasts for 50 seconds. Within each phase, these 13 time series have same frequency but different amplitudes.
The first phase is a sinusoid with frequency $0.4Hz$, in the second phase the sinusoid increases to a frequency of $3.2Hz$, in the
third phase the sinusoid changes to a combination of frequencies $0.4Hz$ and $1.6Hz$,  while in the last phase the sinusoid's frequency shifts to $1.6Hz$.
The sample rate is $400$ samples per second.

For STFT computation we select a window size of 5 seconds, and slide the window every 0.125 seconds.
Due to the simplicity of our signal, we want the STFT to be more precise in the location of the frequency/amplitude transitions.
For topic modeling, we set the number of topics to 3 to match the number of frequencies in the data.
We expect the model to capture the frequencies and separate them into different topics, and the topic transitions should occur approximately when the frequency changes in the signal.

Fig.~\ref{fig:SemanticData} summarizes the results.
As shown in Fig.~\ref{fig:SemanticData}(B), each topic clearly picks out the distinct frequencies in the original data.
Fig.~\ref{fig:SemanticData}(C) shows the time-series document-topic view, where the distinct color changes between topics accurately capture the transitions between frequencies
present in the original time domain. It also accurately splits to two topics when there is a mixture of frequencies in the third phase. The transition between topics along the time series is clearly indicated by the color changes and the time approximately match the frequency changes in the time domain.
This exemplifies the typical use-case of the topic-oriented view STFT-LDA: the user obtains an overview of trends more easily than trying to detect patterns in the original time series.

Fig.\ref{fig:SemanticData}(B) offers an alternative view of the data by emphasizing the topics themselves. The bounding boxes match the colors used in (C). Obviously, each topic identifies one distinct frequency in the signal data.  Since the frequency distribution of each topic is like an impulse function where almost everywhere else  is zero, we expect the topic modeling to pick out the individual frequencies. The opacity of the colors at that frequency encodes the different amplitudes of each time series.

\begin{figure}[!ht]
	\centering 
	\includegraphics[width=\columnwidth]{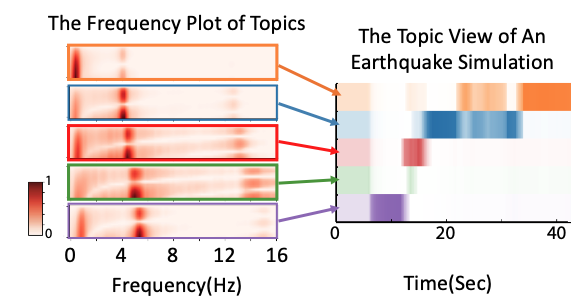}
	\caption{Our topic representation also helps connect to the concept of vibrational modes.  Buildings vibrate mainly in the fundamental natural frequency (first mode) or as damage happens the floors may vibrate out of alignment (second, third modes).  The topic representation helps to see if the floors are vibrating at the same frequencies, which can be verified if the signals are aligned by looking at the signal views.}
	\label{fig:topics}
\end{figure}

\begin{figure*}[t]
	\centering 
	\includegraphics[width=\linewidth]{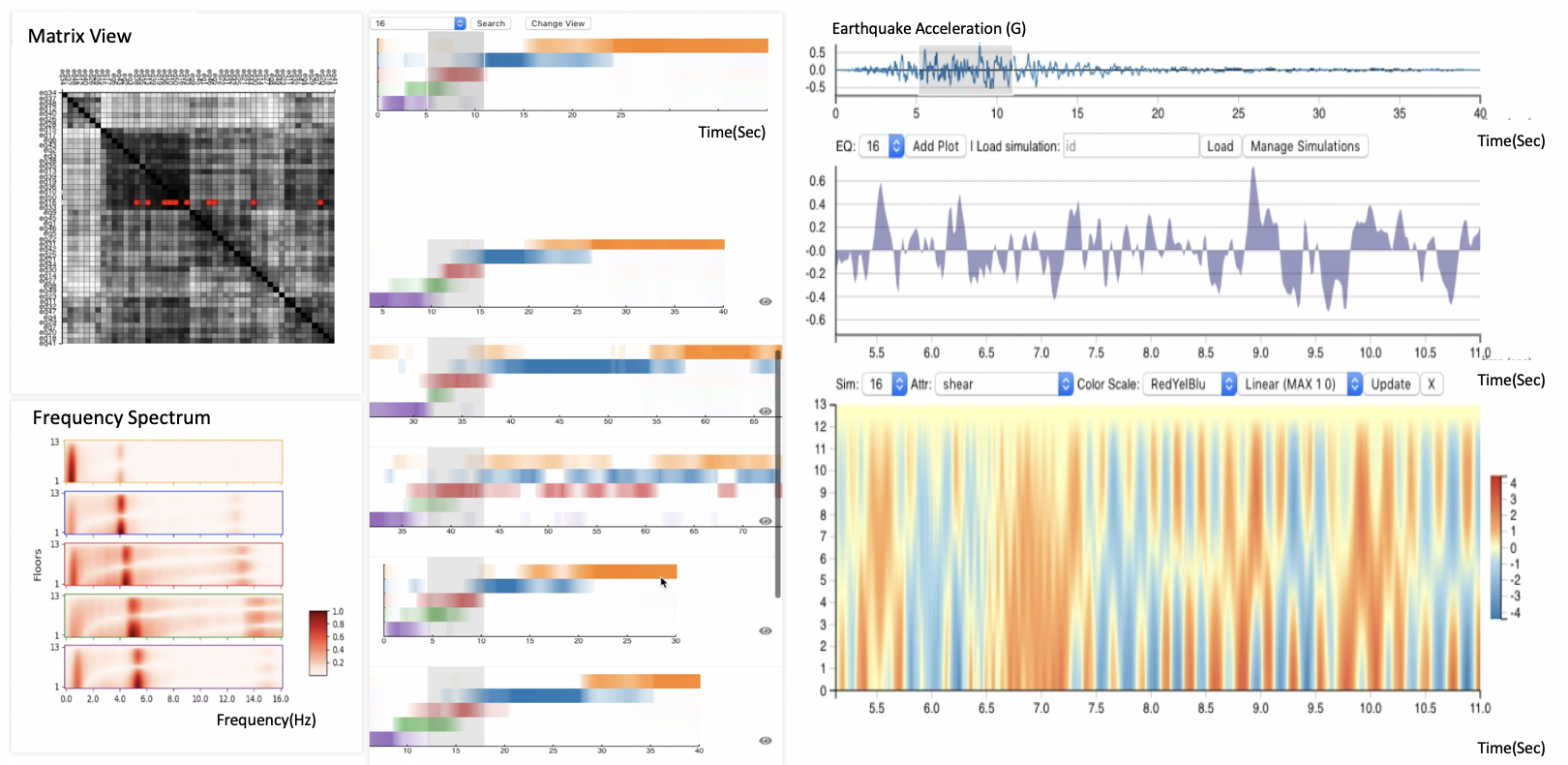} 
	\caption{
		The prototype system consists of four views. The matrix diagram (top left) is 
		used for navigation and summarizes the overall behaviors across all the earthquakes. To understand the frequency distribution of each topic,
		the analysts can refer to the frequency spectrums (bottom left) for details.
		The opacity of the color indicates the relative magnitude of a frequency
		for a specific floor. The core of the system is the topic representation of each earthquake simulation (middle). It includes a content-based search module to help quickly identify similar partial time series across different simulations.
		The last part (right) is a details view that supports further exploration of simulation time series and helps civil engineers interpret the responses of buildings from another aspect.
	}
	\label{fig:system}
\end{figure*}

\section{System Description}
\label{sec:system}

We implemented and experimented with STFT-LDA in a prototype system (Fig.~\ref{fig:system}).
We used an initial collection of 50 simulations of responses to earthquakes. Each building had $13$ floors, and we investigated the variable of shear on each floor, resulting in a $13$-dimensional signal whose lengths varied from 30 to 200 seconds.  Each simulation is the result of a custom simulation developed by two of our coauthors in Matlab where parameters such that the structural method and input earthquake signal were varied. We use python libraries like scikit-learn(https://scikit-learn.org/), NumPy(https://numpy.org/) and SciPy(https://www.scipy.org/) to process the simulation data with different filters like STFT and LDA. We use R for its \texttt{corrplot} package~\citep{corrplot2021} for hierarchical clustering (to reorder the rows and columns of the matrix view) as well as for the analysis of the user study results. All the calculated data are stored in the backend system as binary files in the file system. The application web server is implemented with Flask~\citep{grinberg2018flask}. For the frontend design, we mainly rely on JavaScript library D3~\citep{Bostock:2011:DDD:2068462.2068631} and draw on both SVG and HTML5 Canvas for better performance.
In this section, we discuss each view and interactions we implemented as well as the insights of this simulation dataset we discover by using this prototype system.

\subsection{Topic View}
\label{sec:topic}
To analyze the data, we computed the STFT on each earthquake using SciPy's STFT filter using a window size of 5 seconds with a sampling frequency of 0.125 seconds. The output of the STFT is then processed through Scikit-learn's LDA filter with the batch learning method, setting the number of topics to five.
The visualization for these five topics are shown in Fig.~\ref{fig:topics}(left). The bounding boxes match the color stripes used in Fig.~\ref{fig:topics}(right). Each topic is visualized as a 2D heatmap, where the $x$-axis represents frequencies from $0Hz$ to $16Hz$, the $y$-axis represent 13 floors of the building, and the color opacity indicates the normalized magnitude of each 2D scalar value.
For example, the ``orange" topic has one strong impulse with frequency around $0.4Hz$ and the frequency magnitudes are decaying along the floors. The ``blue" topic picks up a higher frequency of $4.0Hz$, however, the intensities of frequency for each floor diverge from the middle floor of the building. Unlike the first two topics, the remaining three topics contain more mixtures of various frequencies. These five topics summarize common distributions over frequencies and floors across all the simulations. By inspecting these topics, civil engineers can have a better understanding of the general earthquake behaviors. Besides, as the spectrum illustrates the deviations among different floors, our visualization also benefits the analysis of the vibration status of the entire building.

\textbf{Design choices.} In the frequency spectrum, we use a red sequential colormap to indicate the normalized magnitude of the frequency for each floor. We choose five qualitative colors to represents the five topics and we utilize the opacity to represents the percentage of topics at every time interval. 

\subsection{Content-based Search}
\label{sec:content-based-search}

\begin{figure*}
	\centering 
	\includegraphics[width=\linewidth]{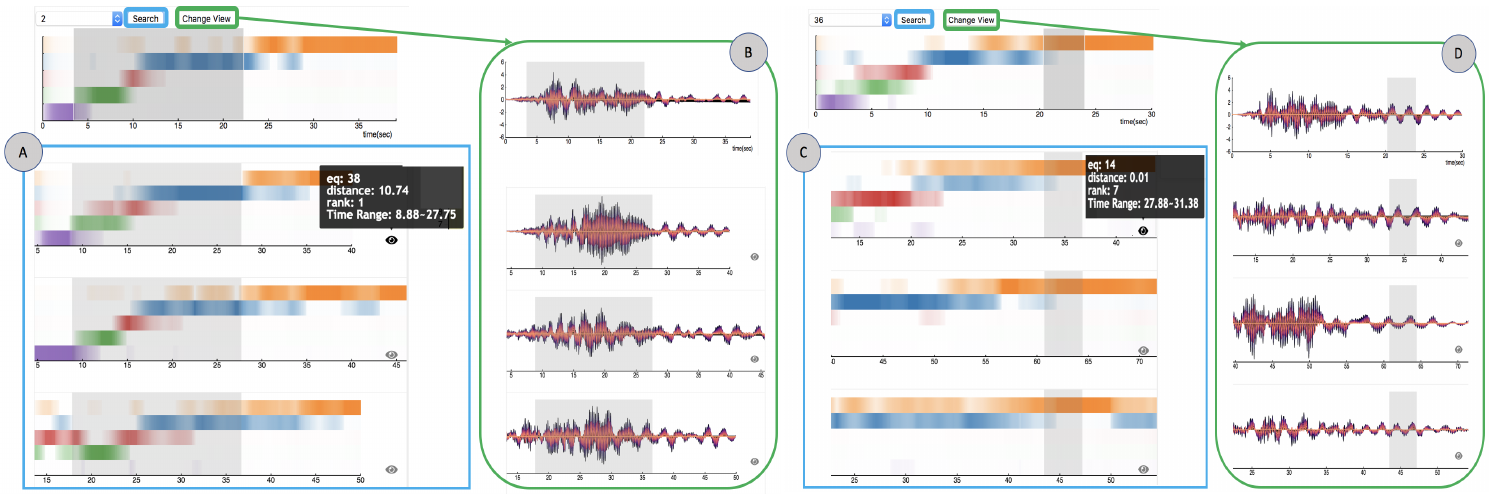} 
	\caption{The content-based search illustrates how topic modeling helps to identify regions that are locally similar and dissimilar.  (A) and (C) show two different brushed regions in the top simulation and three of the most similar results aligned.  For (A), a user can quickly see all three are similar hits and then validate this comparison in (B).  For (C), the topmost hit ends up having a different behavior prior to the selection.}
	\label{fig:search}
\end{figure*}

\zhenge{I have revised this part}
We also implemented an interface for content-based search against temporal regions of earthquakes, see Fig.~\ref{fig:search} for an illustration. This content-based search directly relies on using the topic representations (i.e.~STFT-LDA data). The user can brush on a continuous area of one earthquake simulation and quickly search for the most similar parts of equal length among all the other earthquake simulations, where we use Euclidean distance as the similarity measurement. We then use a cross-correlation process to calculate the \emph{sliding distance} where an accumulation array stores fast Fourier transform to speed up the process (for full details, refer to Appendix~\ref{appendix:search}).  We pick the two most similar parts from each simulation data, order all of them by the distance and return the top simulations. All the search results are translated such that the similar parts are aligned (Fig.~\ref{fig:search}(A,C)). The search results will also be highlighted in the matrix diagram view. We can hover on the icon to show details of the results including the earthquake number, rank, distance between the result part and the search part and the time range of the result part.

This feature also allows the user to compare against the signal view as a validation.  Fig.~\ref{fig:search}(B,D) shows how these two views align.  Shared brushing highlights the same time regions in both views so that users can cross compare.  In particular, this view helps to show both regions where earthquakes are locally similar as well as regions where earthquakes are dissimilar.  

Fig.~\ref{fig:search}(A) shows an example where the user has selected multiple topics and searched for a particular sequence.  The search hits that are returned show three cases that are quite similar.  Fig.~\ref{fig:search}(C) shows a different example, where the user has selected only a single topic to find other earthquakes that express this topic.  As a result, the most similar earthquakes in the selected region of time appear to have a significantly different behavior in the time steps prior to the selection.  The topmost hit (second row) appears to have a more continuous transition between topics, while the next two closest hits (third, fourth row) appear to transition in different ways. The third row shows a more regular transition from the blue to the orange topic, while the fourth row shows that the blue and orange topics appear to be mixed.

\subsection{Other Views}
\label{sec:other-views}
\zhenge{Merge the matrix diagram view and details view to other views}

\subsubsection{Matrix Diagram View}
STFT-LDA splits each earthquake simulation into a set of segments with a fixed window size, and each segment is simply represented as a vector of weights. For any two segments coming from two earthquakes, we compare them directly using the Euclidean distance between these two vectors. Then, we calculate the similarity between any two earthquake simulations by mapping the set of segments to a Gaussian distribution in Hilbert space, and use Bhattacharyya's similarity to compare the earthquakes~\citep{KondorJ03}.

We use matrix diagrams to visualize the behavior across earthquakes as a global comparison mechanism. These are implemented using D3's existing matrix diagram infrastructure. Each cell in the across-earthquake matrix represents the similarity between two entire simulations using Bhattacharyya's measure. In our tool, users can select a cell in the matrix in order to show the details of two earthquake simulations.  In the matrix, we reorder the sequences of the simulations using hierarchical clustering with complete linkage using the R package corrplot.
This matrix view helps to demonstrate how STFT-LDA produces meaningful global
comparisons.  For example, in Fig.~\ref{fig:system} (top left) we can
observe four major clusterings. We can click the corresponding cell to do
pairwise comparisons.  And the the results in the content-based search will also be reflected in the matrix.

\textbf{Design choices.} The matrix diagram is designed as grey-scaled for two reasons: 
\begin{enumerate*}
	\item the sequential color scheme can be used to encode the similarity value;
	\item the color channel can be used for other interactions like highlighting the content-based search results or clicking mark.
\end{enumerate*}

\subsubsection{Details View}
\label{sec:details-views}

\begin{figure*}[!ht]
	\centering 
	\includegraphics[width=\linewidth]{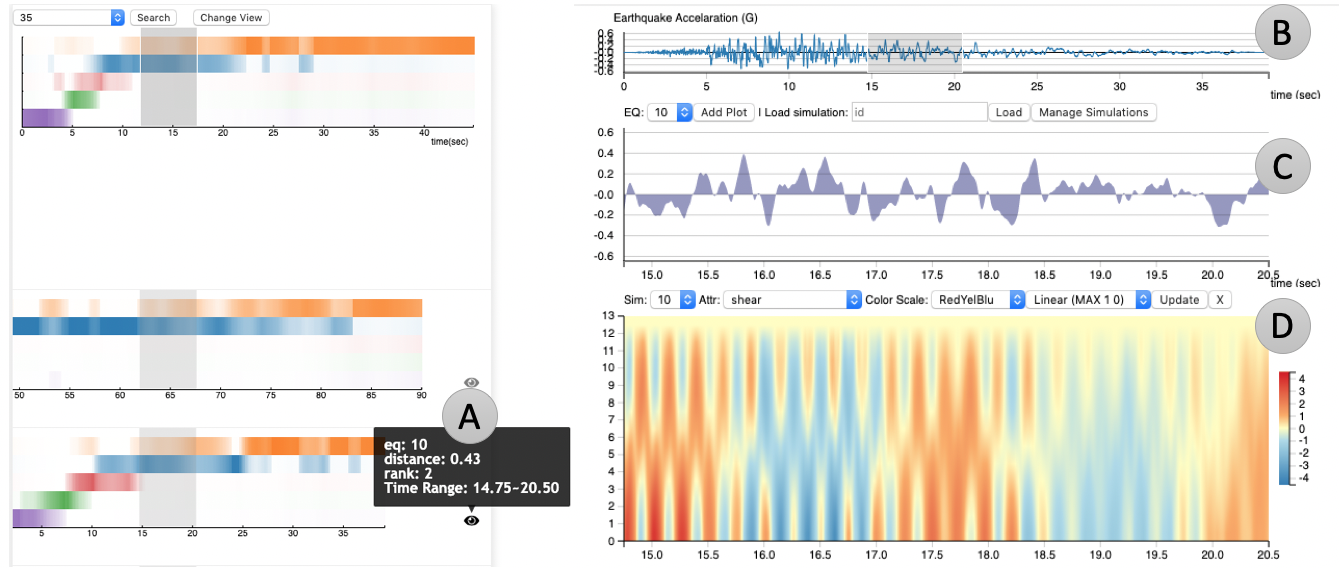} 
	\caption{
		Analysts can select a portion of the ground acceleration (B) and drill down into a specific earthquake simulation (D), to
		visualize the response of a single physical variable plotted over time (x coordinate) and building floor (y coordinate).
		(C) is an area plot visualizing the selected portion of the ground acceleration. By utilizing STFT-LDA, analysts can  quickly navigate and zoom in to the similar partial simulations. In (A) an analyst can quickly switch from \emph{EQ35} to \emph{EQ10} and zoom into time range $14.75s$ to $20.50s$ automatically by clicking on the eye icon. }
	\label{fig:detail}
\end{figure*}

While the topic presentation provides a highly-compressed summary of simulations' overall behaviors, the analysts still prefer a direct visualization of the responses for each floor. This will be a good complement for analyzing the stress condition of different floors. To support further exploration of simulation time series and help civil engineers interpret the responses of buildings from another aspect, the system also includes multiple modules visualizing these time series directly. The views include a line chart visualizing the earthquake acceleration for navigation (Fig. \ref{fig:detail}(B)), an area chart showing the impulses of brushed earthquake time regions (Fig. \ref{fig:detail}(C)), and a 2D heatmap for visualizing the building responses quantified by different physical attributes across all the floors (Fig. \ref{fig:detail}(D)).

\zhenge{justification about why choosing ground acceleration}
Earthquake acceleration is an attribute of the earthquake that directly indicates the intensity the of simulation, and we plot it as a line chart for overview and a gray area plot for details. What's more, we present a 2D heatmap view over time ($x$ coordinate) and building floor ($y$ coordinate) to visualize the response of each single physical variable. User can switch between different simulations and attributes and the positive and negative values of the time series indicate different movement direction of the variable.

We built interactions between the search module and the details view to help 
explore the multivariate data and spot interesting patterns.
For example, as shown in Fig.~\ref{fig:detail}, when a user brushes on a portion of the topic view and searches for similar partial simulations,
these views will automatically zoom into the same time region being searched. What's more, by clicking on the eye icon in the searching results,
user can also quickly switch to highlighted time regions in other earthquake simulations.These interactions enable quick access to the specific time range in earthquakes of the user's interest. They also keep the synchronization of the time range between the time domain and frequency domain and allow the user to analyze similar time intervals discovered by topic views in the time domain as well.

\textbf{Design choices.} We choose diverging color scales for emphasizing the differences and both continuous and discrete colormaps are provided in the view as the former facilitates preserving values and the latter can help filter unimportant values by setting up different thresholds. To help users easily identify patterns over time and floors, we choose to show one attribute each time instead of using small multiples.

\section{Evaluation}
\label{sec:evaluation}

In the previous sections, we argued that STFT-LDA is practical to
implement and provides a number of attractive features in the context of
a larger visual analysis system. However, one central question remains:
\emph{does STFT-LDA actually produce visual representations that
	more readily distinguish different features of the multiple time
	series?} To evaluate the effectiveness of the technique and resulting
time-series visualization, we designed and conducted a surrogate task
with two conditions which we now describe.

\subsection{Surrogate Task}
\label{sec:quantitative-use-study}

Broadly speaking, we sought to study whether participants in the study
would be able to distinguish differences in the features that
generated the time-series data. The true, ecologically-grounded task
of the analyst involves studying, at a potentially fine level,
differences in the behavior of these time series. Such
real-world tasks lack a clear notion of ground truth, making
quantitative experiments particularly challenging. In order to arrive
at one such design, we created a simpler, \emph{surrogate task}, for
which we do have ground truth.

In the study of building responses to earthquakes, engineers create
numerical simulations of a number of different building structures,
and test these structures against the same recordings of earthquakes.
In addition, these simulations have an additional free parameter, the
``load'' of the earthquake, a multiplicative factor of the
ground acceleration that is used to simulate more (or less) severe
versions of the same event.

\begin{figure}[t]
	\centering
	\includegraphics[width=\linewidth]{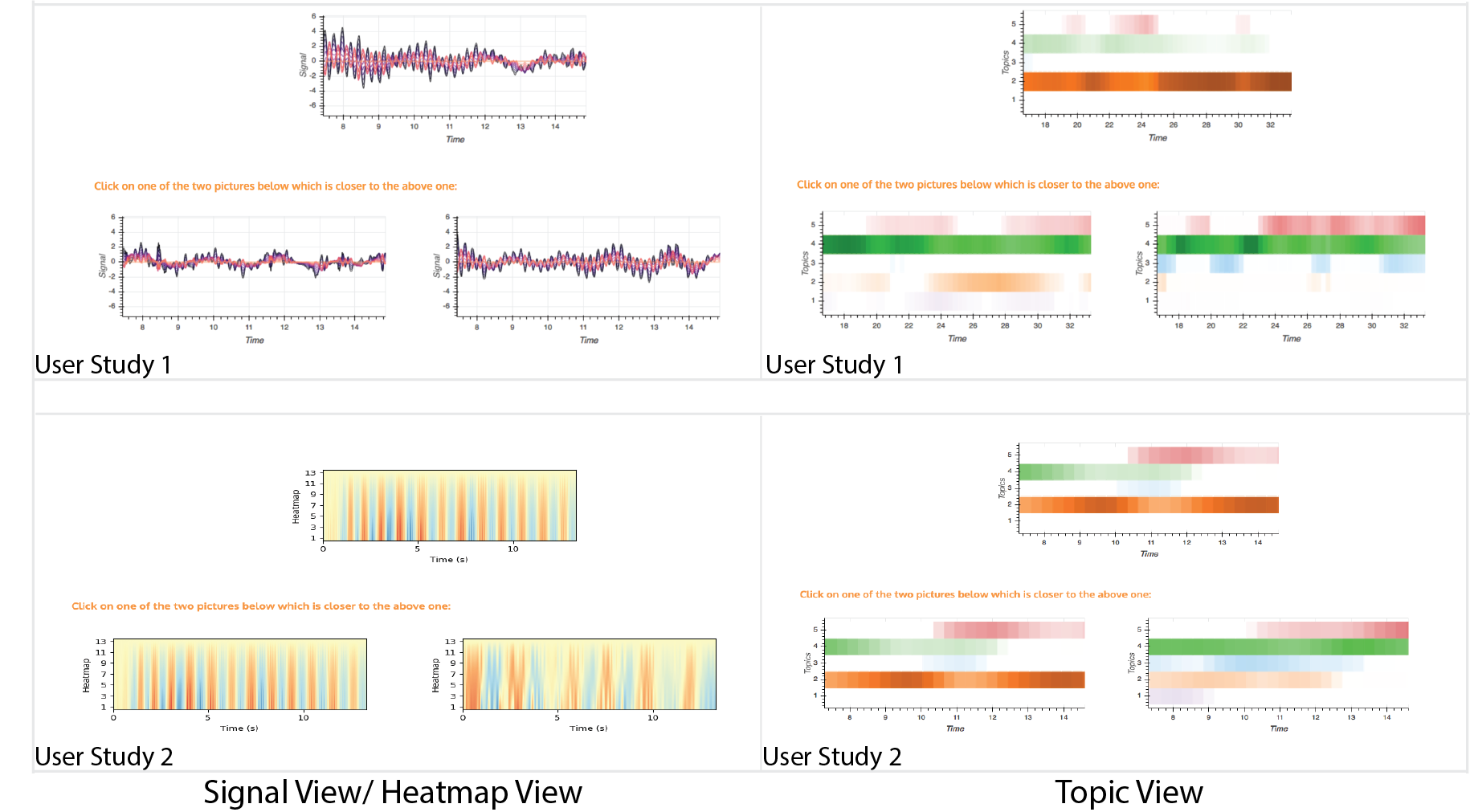}
	\caption{\label{fig:study-stimuli}Some samples from the stimuli
		presented to participants in the surrogate task
		presented in Section~\ref{sec:quantitative-use-study}. The particular
		stimuli presented are chosen to highlight the range of
		variation between easy and hard examples in the surrogate task. The full set of stimuli and source code to reproduce
		the analysis is submitted as supplemental material.}
\end{figure}

The surrogate task we designed is a visualization matching
forced-choice task, where
the participants are shown three stimuli, laid out on a computer
screen as shown in Fig.~\ref{fig:study-stimuli}. Each of the stimuli
shows the same span of time during one fixed earthquake; the
difference in the time series comes from a combination of earthquake
load and building structure. Crucially, one of the images on the
bottom is generated with the same type of building structure as the
one on the top.  Participants are asked to select the image on the
bottom of the screen that looks ``the most similar'' to the one on the
top. We consider the answer correct if it matches the building
structure.

Since ``most similar'' is a markedly subjective notion, and since
participants of the study are not trained in analyzing earthquake
simulation data, we provided a short training session where
participants are given instant feedback as to whether or not they
answered correctly. Although this is not an exactly realistic
scenario, we believe the training session provides information for the
kind of pattern that the analysts should be expected to find in
real-world analyses.

\myparagraph{Hypothesis and Design}
Our hypotheses are:
\begin{itemize}
	\item STFT-LDA will provide higher accuracy in correctly identifying similar patterns, compared to a time-series signal view;
	\item STFT-LDA will provide higher accuracy in correctly identifying similar patterns, compared to a time-series heatmap view.
\end{itemize}
We have done two independent trails for these two conditions. 
For the first trial, the ``visualization'' independent factor is
whether the stimulus is a ``topic view'' (from the results of
STFT-LDA) or a ``signal view'' (from a traditional multiple
time-series view). We use a within-subject design for the
``visualization'' factor, and use randomization to counterbalance the
order in which the factors are presented to each participant. All
participants are shown the same stimuli for the training session,
although the order in which the training session stimuli are presented
is also randomized across participants. The dependent factor in our
study is simply whether or not the participants picked the correct
value, as defined above. Each participant is given a number of such
baseline judgment tasks. 
For the second trail, we keep all the other settings same as
the first one except that the ``signal view" is replaced with a ``heatmap view".

\myparagraph{Pilot study}
We performed an informal, untimed pilot for our study with two participants,
each answering an unlimited number of judgment tasks (until they
informally decided to stop). The exploratory information gathered from
this study suggested we should expect to see around a 10\% absolute
improvement in performance from the signal-based/heatmap-based visualization to the
topic-basic visualization, and we also learned that those participants
did not take more than 10 seconds to answer any of the baseline
judgment tasks. This gave us sufficient information to design an
experiment with sufficient length to give enough power to test the
hypothesis. Ultimately, we arrived at a design where each user answers
30 basic tasks and 4 ``trivial'' tasks designed to exclude
participants who could not understand these instructions. The trivial
tasks showed an identical copy of the target image as one of the
alternatives. We designed the analysis such that if any participant
answers any of the trivial tasks incorrectly, we would discard the
entirety of their input. In addition, the actual responses for the
trivial tasks are discarded.

\myparagraph{Participants}
We recruited a total of 19 participants in the first trial, 
and 22 participants for the second one. 
The time interval between two trails is around 13 months
which minimizes the possibilities of mutual effect between two trials. 
The participants were recruited by local volunteering in classes and
research meetings, and comprise a mix of graduate students and
researchers in computer science and data visualization. Because we
were not interested in post-hoc analysis of demographic information,
we did not formally collect such information as gender or age of
participants. For all the participants, no data was discarded due to
incorrect answers for the trivial tasks.

\begin{figure}[t]
	\includegraphics[width=\linewidth]{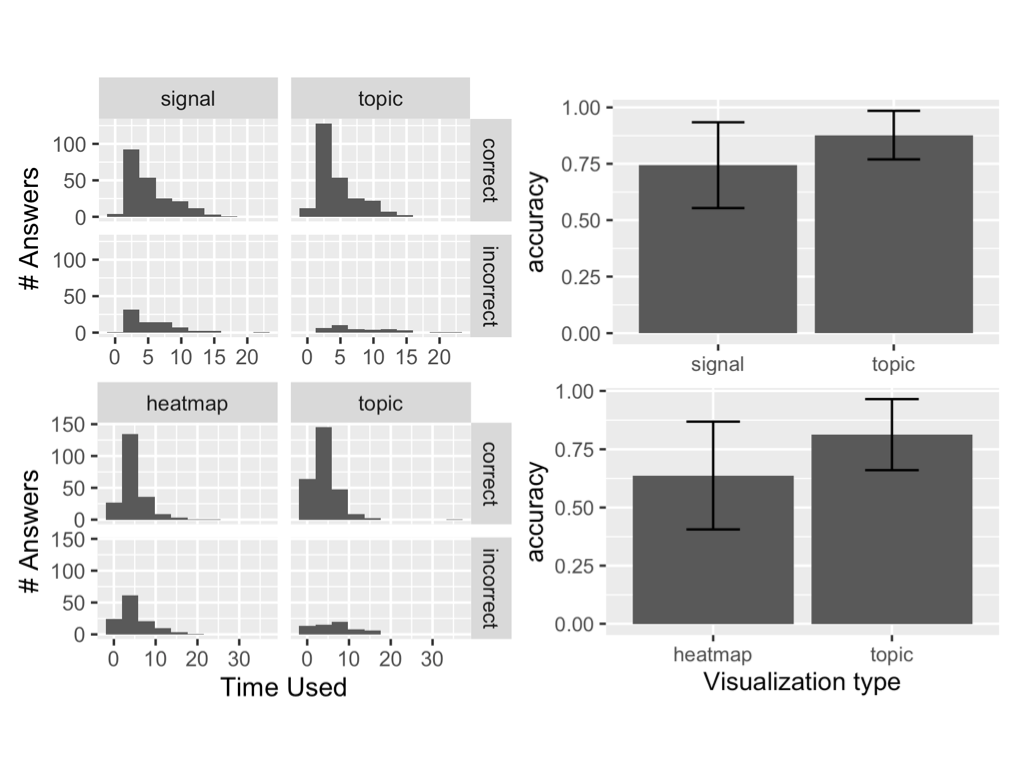}
	\caption{\label{fig:evaluation-analysis} Summary of analysis of surrogate task. On the left, we show a histogram of the times participant took
		to answer the tasks, broken down by whether they answered correctly or
		not, and visualization type. On the right, we show sample accuracy for
		the ``topic'' and ``signal/heatmap" factors, together with the (estimated via
		binomial approximation) standard deviations. We find that the null
		accuracy hypothesis can be rejected with $p=3.36\times 10^{-5}$ and and $p=2.98\times 10^{-7}$, respectively,
		and that at the 95\% confidence level, the odds ratio is smaller than
		$0.61$ and $0.56$, respectively. Relative to the signal-based (heatmap-based) visualization, this means
		participants are likely to be \textbf{more than 60\% (55\%) more accurate in
			the study task using a topic-based visualization}. At the same time,
		we cannot reject the null for the time hypotheses; see the text for
		details.}
\end{figure}

\myparagraph{Analysis}
Our study design enables a relatively simple statistical analysis, in
which we can use Fisher's exact test for count
data~\citep{agresti2003categorical}. The exact count tables for this study
can be seen in Fig.~\ref{fig:evaluation-analysis}. Fisher's exact test allows us
to reject the null hypotheses in two trials at $p=3.36\times 10^{-5}$ and $p=2.98\times 10^{-7}$, respectively, 
and we find that this result is robust under different analysis (which we include
in the supplemental material): an analysis of the odds ratio under the
bootstrap finds similar results, and so does an analysis of the
difference in mean accuracy. We believe this provides adequate
statistical evidence to support our hypotheses. A natural follow-up
hypothesis, then, is: answers using the topic were more accurate because
in those cases users took longer to answer. We test this hypothesis
using a two-sample $t$ test in both two trials, and find that we cannot reject the null,
at $p=0.26$ and $p=0.94$, respectively. In other words, we find strong statistical evidence that
the users were more accurate using the topic-based visual encoding, but no
evidence that they were slower (or faster) using the topic-based
visual encoding.

\myparagraph{Study materials and data}
We have made the study materials, data, and analysis available as part of the supplemental material in the form of CSV files,
R Markdown scripts to reproduce the analysis, and the actual generated
analysis document.

\begin{table*}
	\centering\small\sf
	\caption{\label{tab:questions}
		Questions asked in the expert study. 
	}
	\begin{tabular}{ |c|p{9cm}| } 
		\hline
		TASK & Questions\\
		\hline
		\multirow{2}{*}{T1: Summarizing Earthquake Behaviors} &1. After inspecting each topic heatmap, could you tell us your findings of each topic?  \\ 
		& 2. What is the possible vibration status of the building under each topic?  \\ 
		\hline
		\multirow{2}{*}{T2: Exploring Collection of Earthquakes} &3. By using the matrix view, could you quickly find out two or three similar earthquakes as EQ1 in the dataset? How about EQ12? \\ 
		& 4. Looking at the Matrix View/Scatter plot, how many clusters do you think are there in the dataset?  \\ 
		\hline
		&5. Comparing EQ1 and EQ12 using the topic view, from the frequency and vibration perspective, what do you think are their similarities and differences? How about EQ12 and EQ37?\\ 
		\multirow{2}{*}{T3: Exploring Time-localized Earthquake Features} & 6. In EQ1, we see a topic change from 10s to 20s. Use the search module to find similar patterns in other earthquakes, and tells us what happens from the topic perspective (using the topic heatmap) or from the time series (using the details view)? \\ 
		\hline
		\multirow{2}{*}{T4: Identifying Deviations and Outliers in the Set} &7. Search for one or two earthquakes that are ``unique" in the dataset (different from all the others).  \\ 
		& 8. Why do these earthquakes appear to be unique?  \\ 
		\hline
	\end{tabular}
	
\end{table*}

\subsection{Expert Study} 
\label{sec:expert-user-study}
We conducted an expert study to evaluate the usefulness of our prototype system, where the expert is one of the our coauthors who has worked in structural and earthquake engineering for years.
It was deployed on a cloud server and executed in a web browser (Chrome).
Before the formal study, we piloted the user study with two Ph.D students, who have never seen the visualization.
The information we gathered helped us design each session and estimate time for each question.
Then we scheduled the meeting with the expert via Skype and 
we recorded his computer screen with audio during the conversation. 
The entire user study consists of three parts and took around 90 minutes in total.
The first part is a training session. In this session, 
we quickly introduced STFT-LDA, demonstrated the usage of our prototype system, and
gave the expert enough time to experience the different visualization modules and ask any questions
required to fully understand the visualization approach and its implementation.  
In the second part,  the expert user needs to answer 8 questions using the system (Table~\ref{tab:questions}).
To match the four general tasks we summarize in Section~\ref{sec:task-abstraction},
we designed two questions for each task, 
ranging from summarizing earthquake behaviors to making comparisons between earthquakes.
At last, we collected the expert's comments on both STFT-LDA approach and the prototype system for evaluations and further improvements.

Overall the visualization approach received very positive feedback from the expert.
With the visualization, most of the tasks can be solved much easier than directly utilizing previous visualizations like line plots and our approach significantly reduces the effort in exploring the simulations and benefit the analysis procedure. The user also revealed many interesting findings that were not noticed about the data before.
The results are strongest for Task 1 and Task 3, whereas for Task 2 and Task 4, the expert found the visualizations were a modest help.

The expert agreed that Task 1 is well supported by the visualization approach.
He gave detailed descriptions of the frequency variations for different floors
while inspecting the spectrum of each topic, 
and he immediately associated that with possible vibration modes of the building (Fig.~\ref{fig:topics}).
He also showed strong interests in the search module and visualization of the time series (Fig.~\ref{fig:detail}) and tended to use them together to identify the similar impacts from
different earthquakes and explain these time ranges.
For answering Questions 1 and 2,
the expert analyzed what happened to the building when one particular topic dominates.
For example, based on the topic spectrum,
he speculated that the ``orange topic" is a strong indication of building 
under the \emph{first mode}. 
By brushing on the time range where the ``orange" topic is dominant, 
similar time intervals could be easily found for in other earthquakes.
After inspecting these time intervals in the time series visualization (Fig.~\ref{fig:detail}(D)), 
the expert noticed that all the floors are changing in the same direction while the bottom floor has the strongest vibration intensity.
This confirmed his original guess and he also reported that the earthquake impulses under the ``orange topic'' are often very weak and tend to happen at the end of earthquakes.

The expert also reported that the visualization is helpful in comparing different earthquakes 
in Question 5. He reported that the topic representation simplifies the complicated multi-dimensional time series while preserving discriminative features. 
He stated that he can make quick comparisons by simply identifying
which topic dominates in the earthquake. For further explanations, 
the expert referred to the topic spectrum and vibration modes for details.
This result is also consistent with the conclusion in the quantitative user study (Section \ref{sec:quantitative-use-study}). 

Question 6 is about interpreting the topic transition (``purple-green-red-blue") in Fig.~\ref{fig:search}.
The expert stated that the visualization approach was very helpful in this task.
By utilizing the prototype system, the expert found out that this kind of topic transition indicates the structure damage.
He further explained that when the earthquake first hit the building, the impulse was often small, thus the building was vibrating in a high-frequency, low-intensity \emph{second mode} (``purple topic"); then the magnitude and frequency of the impulse both rose to a very high level(``green topic) and persisted for a small amount of time (``red topic"). Once this large impulse hit the structure, it changed the frequencies of the building due to two reasons: first, the larger impulse caused more serious vibrations; second and most important, this impulse had changed the vibration 
of the building from elastic to inelastic and caused irreversible damage to the structure.
Due to this material damage, the building's vibration changed to a different high-frequency \emph{second mode} (``blue topics''), 
which coordinates with the structure changes. 
The expert stated that finding and analyzing these material damages is one of the main concerns for structural 
engineering and it is supported by the visualization approach.

The lower ratings for Tasks 2 and 4 are somewhat surprising.
The expert stated that although the correlation matrix and the scatter plot provided a summary view for navigation and exploration, 
it was still difficult for him to find obvious clusters or identify outliers in the dataset due to two main reasons: visually, the expert user didn't feel confident distinguishing clusters in the matrix view or associating ``unique'' earthquake with the lightest row or column in the matrix; methodologically, 
instead of showing a single score to indicate the similarity between two earthquakes,
he would like to use the ratios of each topic in an earthquake as a measurement for clustering.

For the modules in the prototype system and interactions between them,
the expert stated that most of the them were very useful. In particular, he was very excited about the search module. Previously, in order to understand how an identified feature in one earthquake relates to other earthquakes, 
the civil engineer needed to manually go through each earthquake.
The expert reported that the search module makes this process much easier
and the response speed is very fast. The expert also appreciated the automatic zoom-in interaction in the details view for brushing in the topic view.
However, the expert user also gave suggestions for improvements.
For example, suggested a view to demonstrate the percentages of each topic within one earthquake. He also expected to see the topic representations of other 
single attribute or multiple attributes. He also suggested ordering the topics by
the absolute energy of each topic.

In summary, the expert agreed that STFT-LDA and the prototype system are very helpful for summarizing responses, especially for identifying similar behaviors across different earthquakes. The expert stated that this method helped solve the problem that the civil engineers can only inspect the data from one perspective, either the frequency distribution, magnitudes or time series for each floor. With our visualization system, the expert could now explore the time series, time window, frequency heatmap, and different behaviors at the same time. 
On the other hand, there were some limitations regarding the identification of clusters and outliers, and for future work we plan to address these issues.

\section{Discussion, Limitations and Conclusion}

The data generated by the earthquake simulations contains, in addition to the shear variable attribute we use in the paper, a number of other attributes such as displacement and moment.
Even though in principle STFT-LDA can handle the summarization of multiple attributes in a natural way, we focus on one attribute for two reasons.
First, it is not clear whether or not a system built to encompass the complete variety of data in the simulation should summarize over these different attributes, or instead provide topic-based views of each of those attributes separately. Second, a quantitative evaluation of the relative merits of these two design decisions is not straightforward. These are both attractive avenues for future work.

STFT-LDA is a practical way of analyzing multivariate time series that combine periodic and non-periodic components. STFT-LDA is capable of capturing the periodic behavior without obfuscating time information. Specifically, it can handle high-dimension data and simplify the complex time series to compositions of different topics. Compared to traditional dimension reduction methods like PCA, the topic components are interpretable and they have specific behaviors. Moreover, it has both flexibility and scalability. For example, as we discussed above analysts can in principle generate topics over two attributes over all floors if they care about the influence of two attributes together. Similarly, STFT-LDA can be used to generate summarizations of all simulation attributes over one particular floor. While a full study of these design decisions is beyond the scope of this paper, extending STFT-LDA to such scenarios is a natural topic for future work.

A particularly promising contribution of the work is exploring time series data through windowed frequency analysis.  While we used the STFT in this work, which we think is particularly amenable to LDA for topic modeling, other approaches such as wavelets or matching pursuit might also be fruitful ways to explore this or similar data.

Our approach has two key user parameters: the window size and the number of topics.  In our work, we used domain knowledge to set the window size to 5 seconds, which was based on domain information of the lowest frequencies of interest that we observed in our simulations.  Using a shorter window would exclude such behaviors, while using a larger window would only increase computation time with no added benefit.  Our work suffers from the same limitation as previous topic modeling works, in that setting the number of topics often requires iteration.  We set the number of topics to five after experimentation.  In our runs, typically, we saw no more expressiveness with using a large number of topics, as new topics typically only captured the transition regions between topics. 

Finally, the expert that we have in Section~\ref{sec:expert-user-study} is also one of our coauthors of this paper. We are aware that having coauthors in the evaluation is potentially problematic, but in this case we lack practical alternatives since they are almost the only people qualified to understand this. Each simulation is designed and run by him and his advisor, as well as the structural method and the input earthquake signal. Thus, they know this unique data better than any other civil engineers. On the other hand, this research is also in the coauthor's thesis~\citep{10150/630560}. They have been investigating relationships between the buildings and the earthquakes for a long time. In summary, to verify if our approach and system can actually help users understand these simulations, they are the most proper domain experts.

In conclusion, we have shown that STFT-LDA is an attractive approach for analyzing periodic and non-periodic features of multivariate time series. Because it accurately captures both local and global features of the time series in a simple descriptor, the visual summaries we can display from the result are more effective at distinguishing seismically relevant characteristics of the simulations. We integrate the full STFT-LDA pipeline into a prototype interactive visualization system. Our surrogate tasks and expert study shows our approach can help civil engineers quickly summarize earthquake behaviors and identify deviations and outliers.

\section{Appendix: Content-based Search}
\label{appendix:search}
Let $\vec{v_1} \in \mathbb{R}^{m \times l}$ be a two dimensional vector, $\vec{v_2} \in \mathbb{R}^{m \times n}$ be another two dimensional vector (assuming $l \leq n$). The rows of two vectors represent the  the temporal range for each topic, while the columns represent the probability distributions of topics for each time step. Given $\vec{v_1}, \vec{v_2}$, the method aims to find a subsequence $\vec{v_3} = \vec{v_2}[:,idx:idx+l-1], 0 \leq idx \leq n-l+1,$ s.t\@.   $\lVert \vec{v_1} - \vec{v_3} \rVert$ is the minimal.

\begin{multicols}{2}
\begin{equation*}
\vec{v_1}=
\begin{bmatrix}
u_{11}& ... &u_{1l} \\
u_{21}& ... &u_{2l}  \\
u_{31}& ... &u_{3l}  \\
...& ... &... \\
u_{m1}& ... &u_{ml} 
\end{bmatrix}
 \end{equation*}
 
\begin{equation*}
\vec{v_2}=
\begin{bmatrix}
v_{11}& ... &v_{1n} \\
v_{21}& ... &v_{2n}  \\
v_{31}& ... &v_{3n}  \\
...& ... &... \\
v_{m1}& ... &v_{mn} 
\end{bmatrix}
\end{equation*}
\end{multicols}

\begin{equation*}
\lVert \vec{v_1} - \vec{v_3} \rVert = \sqrt{(\vec{v_1} - \vec{v_3})^2 } = \sqrt{\lVert \vec{v_1}\rVert^2- 2
	\langle \vec{v_1} \boldsymbol{\cdot} \vec{v_3} \rangle + \lVert \vec{v_3}\rVert^2}
\end{equation*}

We first calculate the cross-correlation, $\vec{a}$, also known as sliding dot product:
\begin{equation*}
\vec{a}=
\begin{bmatrix}
\sum_{i=1}^{m}\sum_{j=k}^{k+l-1}u_{i,j-k+1} \times v_{i,j} \textnormal{ for $k$ from $1$ to $n-l+1$}
\end{bmatrix}
\end{equation*}

Since the cross correlation of two signals is equivalent to multiplication of their Fourier transform:
\begin{equation*}
f \otimes g = F \boldsymbol{\cdot} G
\end{equation*}

A quick way of calculating the cross-correlation is as follows:
\begin{equation*}
\vec{V_1} = fft(\vec{v_1}), \vec{V_2} = fft(\vec{v_2}), \vec{a} =  ifft(\vec{V_1} \boldsymbol{\cdot} \vec{V_2} )
\end{equation*}

We calculate the cumulative sum of $\vec{v_2}$, $\vec{cs}$ as follows:
\begin{equation*}
\vec{cs}=
\begin{bmatrix}
\sum_{i=1}^{m}\sum_{j=1}^{1} v_{ij} ^{2}, \sum_{i=1}^{m}\sum_{j=1}^{2} v_{ij} ^{2} , ... , \\ \sum_{i=1}^{m}\sum_{j=1}^{k} v_{ij} ^{2} , ... , \sum_{i=1}^{m}\sum_{j=1}^{n} v_{ij} ^{2}
\end{bmatrix}
\end{equation*}

Then we can get the array $\vec{b}$:
\begin{equation*}
\vec{b}=
\begin{bmatrix}
\vec{cs}[k+l-1] - \vec{cs}[k] \textnormal{ for $k$ from $0$ to $n-l$}
\end{bmatrix}
\end{equation*}

We also calculate the $l2$ norm of $\vec{v_1}$, and repeat it $n-l+1$ times to get $\vec{c}$:
\begin{equation*}
\vec{c} = [\sum_{i=1}^{m}\sum_{j=1}^{l} u_{ij}^2,...,\sum_{i=1}^{m}\sum_{j=1}^{l} u_{ij}^2]
\end{equation*}

Finally, the full distance array $\vec{d}$ is:
\begin{equation*}
\vec{d} = \sqrt{\vec{c} - 2\times\vec{a} + \vec{b}}
\end{equation*}
We can return the smallest one or two together with the index as the most similar part for a given $\vec{v_1}$.

\myparagraph{Time Complexity Analysis}
An intuitive way for searching a match vector $\vec{v_3}$ from $\vec{v_2}$ for $\vec{v_1}$ needs to calculate Euclidean distance  between two vectors $n-l+1$ times, both two vectors $\in \mathbb{R}^{m \times l}$. The running time is:
\begin{equation*}
T_1  = (n-l+1) \times (m \times l)
\end{equation*}

By using our approach, We can precalculate $\vec{cs}$. Then time for calculating $\vec{b}$ is $n-l+1$. Calculating $l2$ norm of $\vec{v_1}$ takes $m \times l$. Calculating the cross-correlation vector needs two Fast Fourier transform (FFT) and one inverse Fast Fourier transform (Equation 6). Calculating FFT of $\vec{v_1}$, $\vec{v_2}$ takes $(m \times l)\log{l} $, $(m \times n)\log{n} $, respectively. Then calculating the multiplication of $\vec{V_1}$ and $\vec{V_2}$ takes $m \times n$. The final step, calculating the inverse FFT of the multiplication,takes $m \times n \times \log{n}$. In total, the running time is:
\begin{equation*}
T_2 = m \times (l\log{l} + 2n  \log{n} +n )
\end{equation*}

In worst case, when $l = \dfrac{n}{2}$, $T_1 = O(mn^2)$, $T_2 = O(mn\log{n})$, the difference is significant.
We run experiments using both methods on the current dataset, it turns out, the average search time using the intuitive way is $168.8ms$, but using our method, the average search time is around $33.5ms$.

\bibliographystyle{SageH}
\bibliography{ms.bib}

\end{document}